# Electric Field Properties inside Central Gap of Dipole Micro/Nano Antennas Operating at 30 THz

**Liangrid Lutiani Silva[1,2]\*, Newton Adriano dos Santos Gomes[1], and Vilson Rosa Almeida[1,2]**

[1]*Instituto Tecnológico de Aeronáutica, São José dos Campos - SP, Brazil*
[2]*Instituto de Estudos Avançados, São José dos Campos - SP, Brazil*

*\*Corresponding author: liangrid@ita.br*

*This work investigates the influence of geometric variations in dipole micro/nano antennas, regarding their implications on the characteristics of the electric field inside the gap space of antenna monopoles. The gap is the interface for a metal-Insulator-Metal (MIM) rectifier diode and it needs to be carefully optimized, in order to allow better electric current generation by tunneling current mechanisms. The arrangement (antenna + diode or rectenna) was designed to operate around 30 Terahertz (THz).*

## Introduction

Micro and nano-antennas have allowed new perspectives in micro/nanotechnologies, especially in the photonic scenario. The possibility to advance in THz spectral domain [1], [2], as well as to explore subwavelength applications as single photon emitters [3], sensors [4], intrachip communications for advanced interconnection communication data meshes in integrated circuits (IC) designed to manipulate packages of information in optical and electrical domains concomitantly [5], [6], amongst others.

On the energy harvesting and detection realm, the rectennas schemes have been extensively studied, not only to operate in the infrared spectrum [7], but also in optical frequency regimes [8]. The physical process to catch a free space electromagnetic wave is analogous to that of RF antennas counterpart. However, one of the main points to consider in rectenna design is the interface between antenna-MIM diode (Metal-Insulator-Metal diode) due to electric current generation (d.c.) which is dominated by the current tunneling mechanisms (CTM) for fast operations [2], [9], [10]. Additionally, the plasmonic research field have been studied and developed in order to overcome the antenna-scaling limits, as well as to improve the light confinement [11], [12], and this approach has extended from optical to infrared spectrum [2], [13].

The objective of this work is to analyze the behavior of the electric field confined between the gap region (i.e. barrier thickness between Al-Cu interfaces in MIM assumptions, filled by atmospheric air) of two distinct dipole antennas, in order to provide ways to attain better performance for tunneling current generation, as well as to evaluate the characteristics of THz confinement. The antennas were designed for around 30 THz ($\lambda \approx 10$ μm). We highlight that the tunnel current mechanism models are not explicitly considered in this work [14]. In the next section, we present the materials and methods adopted to develop this study. The results are shown in the following section and, finally, our main conclusions are presented.

## Materials and Methods

One of the dipole antennas has a straight edge termination (Fig.1-a), whereas the other one has a round termination (Fig.1-b). By means of simulation analysis, the effect of geometric (width) variation (parameter $Lm_1 \in \{300, 250, 200, 150\}$ nm) of the variable arm of each antenna were evaluated. Aluminum (Al) was adopted as the material of the variable dipole antenna arm, whereas Copper (Cu) was adopted to the fixed arm., and atmospheric air was considered as the insulator material. This configuration was chosen by convenience, in order to allow an asymmetric metal-insulator-metal arrangement. The antennas were simulated on SOI substrate, where the thicknesses used were: 2.3 μm, 380 μm and 300 nm, respectively, for $SiO_2$, Si and metallization (Al) (see fig.1-c).

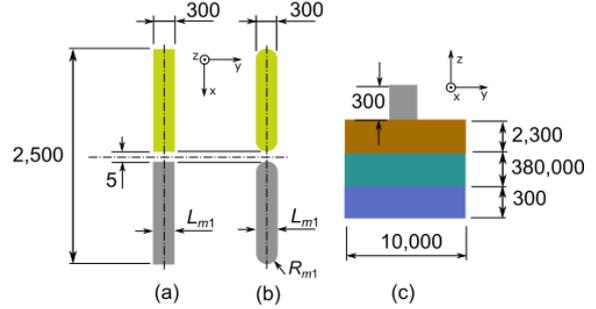

Fig. 1 – Schematic view of dipole antennas. (a) Straight face gap; (b) radius face gap; (c) cross section view. All dimensions are in nanometer unit.

The optical material properties of Si and $SiO_2$ insulators were modeled according to the experimental data collection performed by Palik [15] and implemented as

$$\varepsilon_r(\omega) = \varepsilon_{r1}(\omega) + i\varepsilon_{r2}(\omega), \quad (1)$$

where $\varepsilon_r(\omega)$ is the complex relative dielectric function, $\varepsilon_{r1}(\omega)$ and $\varepsilon_{r2}(\omega)$ are, respectively, the real and imaginary components of dielectric function. In addition, the relative dielectric function of metals were modeled according Lorentz-Drude model based in Johnson and Christry [16], and Rakić and Marković [17], [18] experimental fitted data for Cu and Al materials, respectively and it is given by [18]:

$$\varepsilon_{r1}(\omega) = 1 - \frac{\omega_p^{\,2}}{\omega^2 + \Gamma^2} \quad (2)$$

$$\varepsilon_{r2}(\omega) = \frac{\omega_p^{\,2}\Gamma}{\omega(\omega^2 + \Gamma^2)} \quad (3)$$

where $\omega$ is the frequency of incident light, $\omega_p$ is the plasma frequency and $\Gamma$ is the damping frequency. The study simulations were performed with a commercial full electromagnetic solver (HFSS™), by means of finite element method approach.

## Results

Figure 2 shows the results of electric field profile along the antennas longitudinal gap length, which were measured in the center of x-z plane (see Fig. 1) of each antenna for all simulated cases. It is evident for all cases that the dipole antenna arms ended by radius edge presents stronger amplitude of their electric fields, as compared with the antennas with straight edge termination.

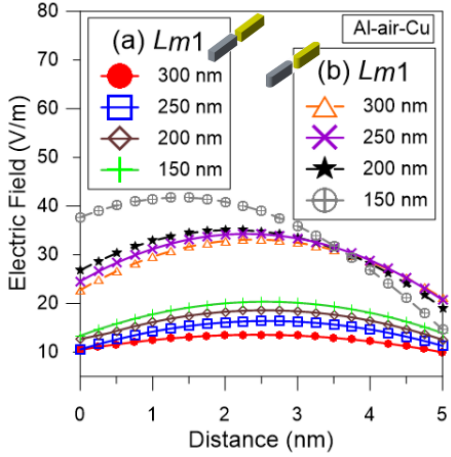

Fig. 2 – Electric field amplitude along dipole antennas gap distance, varying the $L_{m1}$ parameter.

The reason for this can be explained due to the increase of charge density of antenna tips, by means of Gauss divergence. The antenna with radius edge termination with $L_{m1}$ = 150 nm presents the maximum value of electric field with 42.3 V/m. Additionally, it was observed that most of the electric field curve responses present a slow asymmetry in their electric field distribution. The asymmetric electric field ratio (AEFR) on gap was defined as $E_{g0}/E_{g5}$ where, $E_{g0}$ is the magnitude of electric field at distance value equal to 0 nm (i.e. the border side of Al material) and $E_{g5}$ is the magnitude of electric field at distance value equal to 5 nm (i.e. the border side of Cu material) (see Fig. 1). The AEFR results are shown at Table 1. First, the dipole straight edge shows the ratio with smooth increase of variation when $L_{m1}$ range decreases. On the other hand, for dipole radius edge, the AEFR increases fast when the $L_{m1}$ range decreases. The results highlight the effects of both dissimilar metals chosen for each dipole arms, as well as the high field confinement supported by radius tips, which corroborates with Gauss law.

Table 1 – Asymmetric electric field ratio on gap ($E_{g0}/E_{g5}$)

| $L_{m1}$ (nm) | Straight Edge (a) | Radius Edge (b) |
|---|---|---|
| 300 | 1.03 | 1.08 |
| 250 | 1.03 | 1.18 |
| 200 | 1.05 | 1.41 |
| 150 | 1.06 | 2.58 |

Figures 3 and 4 show the magnitudes of electric fields inside the gap, on the x-y plane, which is highlighted by red dash lines with dimensions of 300 nm × 300 nm centered in the gap region, for each dipole antenna over the $L_{m1}$ range values (see the center of Fig. 3 and Fig. 4, respectively).

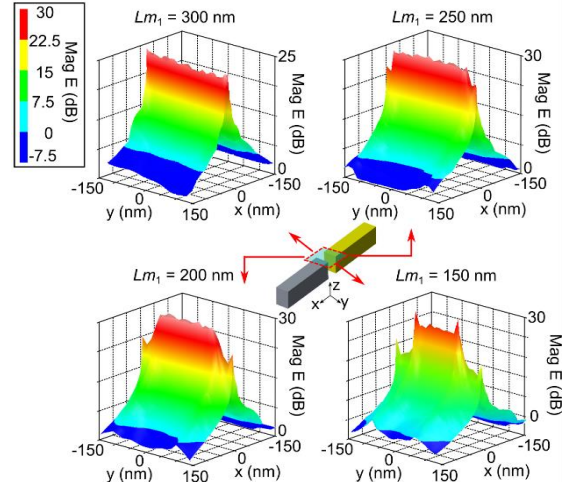

Fig. 3 – Magnitude of electric field along straight-edge dipole antennas gap distance varying the $L_{m1}$ parameter. These fields were measured on the x-y plane which is highlighted by red dash lines, with dimensions equal 300 × 300 nm and centered in the gap region.

For dipole antenna with straight edge termination, the spatial electric field confinement is evident, and it shows an increase when the $L_{m1}$ values decrease. These complementary plots highlight the most pronounced electric field tips or edge effects, which are not captured in the previous analysis. In the $L_{m1}$ plots from Fig. 3, these edge effects are weakly pronounced in terms of electric field magnitude for $L_{m1} \in \{300, 250, 200\}$ nm. However, for $L_{m1}$ = 150 nm, the edge effect is significantly more pronounced. For dipole antenna with round-edge termination, the electric field confinement is enhanced as the $L_{m1}$ values decrease (see Fig. 4). It is evident that the edge effect occurs only in a tangent point of the radius.

Based on the results of electric field confined on gap region of each type of dipole antenna, we can make some qualitative assessments about the current mechanism process. Firstly, according to the most acceptable CTM models, there are several parameters which need to be known, in order to improve the model and results, such as work function of metals, dielectric permittivity, electric field intensity, interfacial potential barriers and so on [14].

Considering the electric fields only, for gap values equal or smaller than 5 nm, the CTM model approach commonly applied is based on Schrödinger equation derivation [19], [20]. However, when the electric field is high and the barrier thickness has the same characteristics described above, the dominant conduction regime is the Fowler-Nordheim tunnel effect [21], [10], [22]. In this way, based on our simulation results, we are encouraged to highlight the importance to verify the fields distributions, as well as their respective magnitudes, specially if other materials are selected as insulators for the gap region; such considerations may be relevant to choose the best CTM model for optimized estimative of current flow through the interfaces. Afterwards, based on the electric field evaluation, the influence of edge effects need to be carefully analyzed, in order to determine whether the edge effects behavior will be taken into account in CTM models or not.

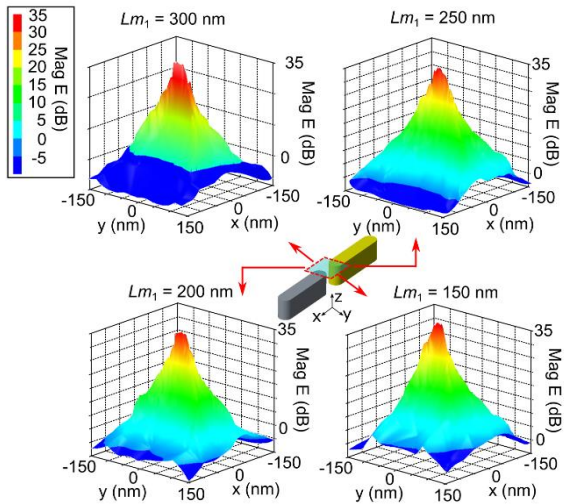

Fig. 4 – Magnitude of electric field along round-edge dipole antennas gap distance varying the $L_{m1}$ parameter. These fields were measured on the x-y plane which is highlighted by red dash lines, with dimensions equal 300 × 300 nm and centered in the gap region.

It is important to highlight that, in a viewpoint of actual fabrication processes technology, the straight edge termination of antenna gap is a very hard or unpractical task to achieve, especially when the smallest dimensions are reduced below 100 nm, in order to tune the antenna geometry to operate in near IR and optical frequency regimes. In most cases, the intrinsic defects are related to lithographic mask resolution and cannot be extinct, although they can be mitigated [23].

## Conclusions

Electromagnetic simulations were employed to investigate the dependence of electric field behavior in the dipole antenna's gap region. Our main conclusions point out that the antenna's end edge shape, as well as the width of monopole arms, for each antenna model presented, need to be carefully adjusted, in order to improve the electric field amplitude and confinement. In addition, we highlight that some tip border effects may be relevant in the current tunnel models, which may demand some adjusts in a common one-dimensional current tunnel model approach [10].

This work results are useful for optimization and maximization of charge transport mechanisms, by means of tunneling current models used in MIM diodes designs, which will be object of future quantitative studies, in order to provide better evaluation of electric current generation supported by rectenna systems.

## Acknowledgment

This work was partially funded by CAPES and CNPq. L.L.S. thanks CAPES for the scholarship, and V.R.A. thanks CNPq grant n: 310855/2016-0 and CAPES for PVS-CAPES/ITA n. 48/2014.